\documentclass[aps, prx, notitlepage, twocolumn,longbibliography,nofootinbib,floatfix]{revtex4-2}

\usepackage{amssymb}
\usepackage{amsmath}
\usepackage[colorlinks=true]{hyperref}
\usepackage[caption=false]{subfig}
\usepackage{tabularx}
\usepackage{cancel}
\usepackage{comment}
\usepackage{algorithm}
\usepackage{algpseudocode}

\usepackage[utf8]{inputenc}
\usepackage{color}
\usepackage{graphicx}
\usepackage{subfig}
\usepackage{empheq}

\newcommand{\bra}[1]{\left\langle#1\right|}
\newcommand{\ket}[1]{\left|#1\right\rangle}

\def\bea{\begin{eqnarray}}
\def\eea{\end{eqnarray}}

\def\ba{\begin{array}}
\def\ea{\end{array}}
\def\Tr{\text{Tr}}

\usepackage[shortlabels]{enumitem}

\begin{document}

\date{\today}
\title{Efficient tensor network simulation of quantum many-body physics on sparse graphs}

\author{Subhayan Sahu}
\email{subhayan@terpmail.umd.edu}
\affiliation{Condensed Matter Theory Center and Department of Physics, University of Maryland, College Park, MD 20742, USA}
\author{Brian Swingle}
\affiliation{Martin A. Fisher School of Physics, Brandeis University, Waltham MA, USA}
\affiliation{Condensed Matter Theory Center and Department of Physics, University of Maryland, College Park, MD 20742, USA}

\begin{abstract}
We study tensor network states defined on an underlying graph which is sparsely connected. Generic sparse graphs are expander graphs with a high probability, and one can represent volume law entangled states efficiently with only polynomial resources. We find that message-passing inference algorithms such as belief propagation can lead to efficient computation of local expectation values for a class of tensor network states defined on sparse graphs. As applications, we study local properties of square root states, graph states, and also employ this method to variationally prepare ground states of gapped Hamiltonians defined on generic graphs. Using the variational method we study the phase diagram of the transverse field quantum Ising model defined on sparse expander graphs.
\end{abstract}

\maketitle

\section{Introduction}\label{sec:intro}
Quantum many-body physics is generally studied on regular d-dimensional lattices since the underlying graph is motivated by naturally occurring crystalline solid state materials and lattice regularizations of quantum field theories. However, there are interesting quantum phenomena beyond those feasible on lattices. On one hand, properties of most topological phases of matter do not depend on the exact triangulation of the underlying manifold. In the other extreme, one can study interacting quantum many-body systems on underlying graphs which do not have any smooth manifold structure. The corresponding classical problem was first studied by Bethe~\cite{bethe_statistical_1935} in the context of alloys, and since then in the context of spin glasses on random graphs~\cite{Mezard_2001}. Previous studies of quantum many-body systems focusing on Bethe lattices and generic sparse graphs have identified several interesting phenomena including approximate solvability leading to mean-field numerical methods~\cite{Georges_RMP_1996}, quantum spin glass states~\cite{Laumann_2010}, and the absence of Goldstone bosons on `expander' graphs~\cite{Laumann_2009}. 

From a modern quantum information perspective, many-body sparse graphical models typically possess the feature of fast quantum information scrambling, which was demonstrated first in all-to-all connected graphical models such as the Sachdev-Ye-Kitaev (SYK) model \cite{kitaev2015,Sachdev_2015}. These models are holographically dual to a quantum theory of gravity in one higher dimension~\cite{kitaev2015,maldacena2016remarks,Sachdev_2015}. In a SYK-like model on $N$ sites, any local quantum information spreads across the whole system in a short scrambling time, $t_{*}\sim \log N$~\cite{Hayden_2007,lashkari2013towards}. On the other hand, in generic local models on d-dimensional lattices, typical scrambling times are long $t_{*}\sim N^{1/d}$. In fact, having a complete (i.e. all-to-all connected) graph is not necessary for getting fast scrambling - generic sparse graphs can also scramble information quickly while retaining the feature of approximate solvability \cite{bentsen2019fast,xu2020sparse}. Sparse graphical models are also attractive platforms to be simulated on a quantum processor, since the sparse connectivity of the graph can lead to efficient quantum simulation. There have already been efforts to realize non-trivial graphs as the platform for many-body physics in quantum simulation architectures \cite{Kollar_2019}. In this context, reliable classical algorithms to simulate quantum many-body models on sparse graphs are highly desirable.

\begin{figure}
\centering
\includegraphics[width = 0.9\columnwidth]{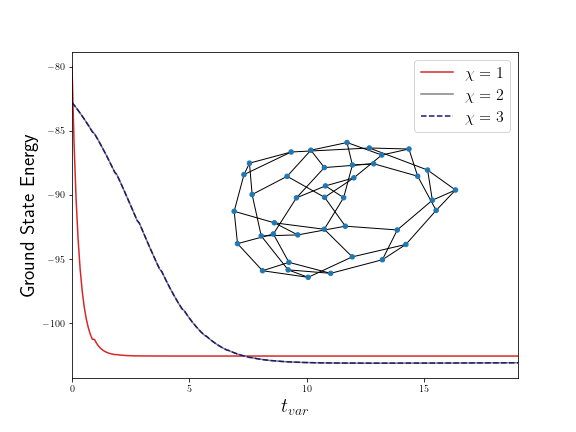}
\caption{\textbf{Approximate ground state preparation} for a mixed-field Ising model defined on a random regular graph on 40 vertices [inset]. The parameters of the Hamiltonian for the local terms coupling the quantum spins on nearest neighbors on the graph are $J_{zz}= - 1$, and on-site terms $h_{x} = -2, h_{z} = -0.5$. The variational algorithm is described in Sec.~\ref{sec:variational_prep}. Here we show that the ground state energy has converged by increasing the bond dimension $\chi$ from $1-3$.}
\label{fig:variational_gs_prep}
\includegraphics[width = 0.9\columnwidth]{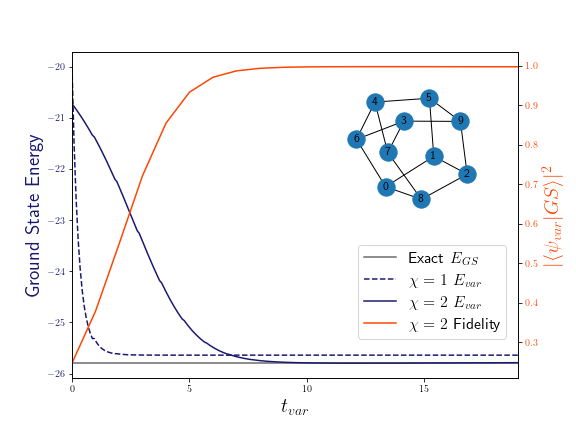}
\caption{\textbf{Approximate ground state preparation} for a mixed-field Ising model defined on a random regular graph on 10 vertices [inset]. The parameters of the Hamiltonian are same as Fig.~\ref{fig:variational_gs_prep}. Here we show that the ground state energy of the variational ground state $\chi = 2$  has converged to the exact ground state energy value, which is accessed by exact diagonalization. We also estimate the overlap of the variationally prepared state with the exact ground state obtained from exact diagonalization, $|\bra{\psi_{var}}GS\rangle|^{2}$, which goes to $1$ after a few steps of the variational algorithm.}
\label{fig:variational_gs_prep_benchmark}
\end{figure}

Tensor network states are useful classical ans\"atze for representing and manipulating entangled quantum states \cite{Cirac_2021,Orus_2014}. They have been used to study quantum many-body systems, most successfully in 1 spatial dimension, where tensor networks are routinely used to study ground state properties of gapped Hamiltonians \cite{Affleck_1987,White_1992}, and simulate short-time quantum time evolution \cite{Vidal_2003}. 

Tensor network states, more specifically matrix product states, can be readily generalized to higher dimensions \cite{Verstraete_2004}. In higher dimensions the representation of such states is still efficient, i.e. the numerical resources required to represent the state scales polynomially with the number of sites of the underlying graph. However the computation of expectation values of local operators is prohibitively hard; one dimension is special because there exists an efficient way to contract tensor networks that fails in higher dimensions. In fact, contraction of generic tensor network states in a 2-d lattice or Projected Entangled Pair States (PEPS) is $\#P$ complete~\cite{Schuch2007}, so any NP hard problem can be encoded in such tensor networks. However, approximate methods of contracting classes of 2-d tensor networks (for example, \cite{Verstraete_2004,Murg_2007,Jiang_2008}) or efficient manipulation methods for a restricted class of 2-d tensor networks (for example, \cite{Zaletel_2020,vanderstraeten2021variational}) are still very useful.

In this work, we demonstrate that approximate local properties of certain class of tensor network states defined on sparse graphs can be efficiently computed using message passing or Belief Propagation (BP) algorithms ~\cite{pearl_fusion_1986}. Efficient classical simulation of generic sparse graphical models are severely restricted by the presence of cycles or loops in the graph. However, we show in this work that the locally tree-like property of generic sparse graphs allows us to efficiently study the properties of quantum states on such graphs using tensor network contraction by belief propagation. Before describing the details, we first demonstrate the usefulness of such a method. In Fig.~\ref{fig:variational_gs_prep}, we variationally access the ground state properties of strongly interacting mixed-field quantum Ising model defined on a sparse random regular graph with $40$ vertices. These system sizes are inaccessible to exact diagonalization, and traditional tensor network methods also do not work well for such graphs, as tensor network contraction is severely affected by the presence of cycles in the graph. However, we show in this work that the BP algorithm can efficiently compute local energy functionals for tensor network states defined on such graphs, which allows us to systematically access the ground state energy in a standard laptop in a few minutes. In Fig.~\ref{fig:variational_gs_prep_benchmark} we show by comparison to an exact computation in a graph with 10 vertices that the prepared tensor network state is indeed the ground state of the Hamiltonian considered. In
later sections we will describe the method and demonstrate careful numerical benchmarks.

A typical class of problems where BP can be employed is in extracting marginal distributions from Gibbs distribution of classical spin models. Consider a classical spin model defined on a graph $G$, with the Hamiltonian $H = -\sum_{a, b\in \mathcal{N}_{a}}h_{ab}(s_{a},s_{b})$ where $\mathcal{N}_{a}$ refers to the graph neighborhood of the site $a$, and $h_{ab}$ refers to the local energy on an edge $ab$. The Gibbs distribution $P(\{s_{i}\}) \propto e^{-\beta H(\{s_{i}\})}$ is efficiently represented by specifying the edge energy functionals $h_{ab}$ for all edge $ab \in G$ that connect the vertices $a$ and $b$. However, accessing marginal probability distributions of one (or few spins) requires one to contract the Gibbs distribution over the graph $G$, which can be hard, and BP algorithms can provide an approximate solution to this problem. A generic theory for the success of BP is still an area of active research, however, it has been shown that the results obtained from BP algorithms are equivalent to the Bethe Peierls approximation (where the underlying graph is assumed to be an infinite tree Bethe lattice)~\cite{yedidia2003understanding}. On a tree graph (which by definition lacks cycles), the BP algorithm is exact, while on a graph with cycles there is no guarantee that the algorithm will converge or provide the correct answer. However, these algorithms are routinely used even for graphs with cycles, and empirically provide correct answers when the underlying graph is locally tree-like~\cite{kschischang2001factor,mezard_information_2009}. In fact, BP has been instrumental in decoding classical low-density parity-check (LDPC) error correcting codes~\cite{mezard_information_2009}.

Tensor network states can be mapped to the graphical models described above~\cite{Robeva_2017}, where the amplitude corresponding to the tensor network state is analogous to the Gibbs distribution over a classical graphical model, and accessing the local reduced density matrices is analogous to marginalizing the Gibbs distribution. Based on this understanding, it was shown by Leifer and Poulin that tensor network contraction can be done using BP~\cite{Leifer_2008}. More recently, Alkabetz and Arad~\cite{Alkabetz_2021} showed that BP algorithms can be used to access local observables in PEPS defined over 2-d lattices, and provide the same answer as other widely used approximate PEPS contraction methods~\cite{Jiang_2008}. Led by the intuition of BP being more successful when the underlying graph is locally tree-like, we employ BP to study tensor network states on sparse locally tree-like graphs, and develop variational methods to prepare tensor network states that are approximate ground states of local Hamiltonians defined on such graphs. This also allows us to study interesting physics questions such as the phase diagram of a transverse field Ising model across the symmetry-breaking quantum phase transition. 

Local properties of thermal states of sparse graphical Hamiltonians have been previously studied using the quantum cavity and quantum belief propagation methods~\cite{Cesare_1992,Hastings_2007,Poulin_2008,Laumann_2008,Krzakala_2008}, which are quantum formulations of the BP-inspired classical cavity algorithms~\cite{Ferraro_2014}. The idea in those works is to represent the quantum partition function as a classical probability distribution and find its marginals using belief propagation. Our work builds a bridge between those methods and the problem of tensor network contraction on generic graphs, which can lead to future cross fertilization of these fields.



Let us briefly comment on the layout of the rest of the paper. We first introduce notation and brief definitions for graphs and sparse graph tensor network states in Sec.~\ref{sec:graph_tn}. In Sec.~\ref{sec:bp} we describe the BP algorithm for contracting tensors and explain the intuition why the method is expected to work for  accessing local expectation values for sparse tensor network states. In Sec.~\ref{sec:graph_sqroot} we demonstrate the viability of this method by computing local operator expectation values for a variety of graph-like quantum states on random regular graphs. Finally, in Sec.~\ref{sec:variational_prep} we use the BP contraction method to variationally prepare ground states of local Hamiltonians defined on sparse graphs. This allows us to study the phase diagram of a transverse field Ising model on a random regular graph across the usual $\mathbb{Z}_{2}$ symmetry-breaking transition. We end by commenting on the prospects of the BP contraction methods in tensor networks, and studying many-body physics on sparse graphs.

\section{Tensor networks on graphs}\label{sec:graph_tn}
\begin{figure}
    \centering
    \includegraphics[width = \columnwidth]{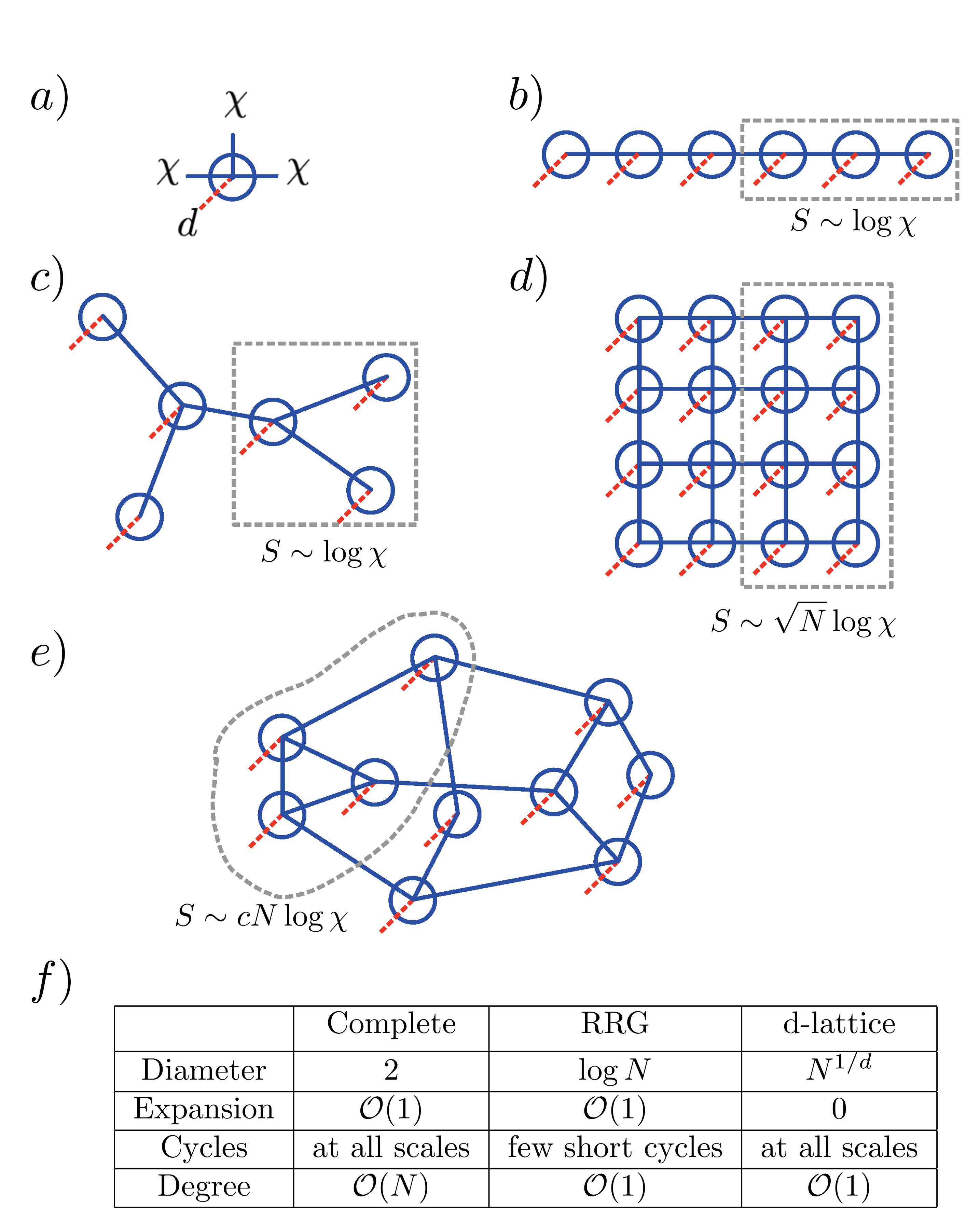}
    \caption{\textbf{Tensor networks on generic graphs.} The fundamental component is the on-site tensor \textbf{(a)}, with physical dimension $d$ and virtual bond dimension $\chi$. These can be put on any underlying graph: 1-d lattice \textbf{(b)}, a tree \textbf{(c)}, 2-d lattice \textbf{(d)}, and a random regular graph(RRG) $\textbf{(e)}$. We also show the scaling of the maximal entanglement of the tensor network ansatz for a typical fraction of the graph \textbf{(b-e)}. On RRG, volume law states can be represented by finite $\chi$ tensor networks. In \textbf{(f)} we compare the graph properties of a complete graph, random regular graph (RRG) and a d-dimensional lattice. The properties being compared are: diameter or maximal distance between any two vertices, expansion defined in Eq.~\ref{eqref:expansion}, number and type of cycles, and degree or number of neighbors of any vertex.}
    \label{fig:tn_graph}
\end{figure}

A graph $G(V,E)$ is specified by its set of vertices $V$, and the set of edges $E$ connecting any two vertices. $G$ is $r-$regular if the degree, or the number of neighbors of each site, is constant and equals to $r$. If any subgraph of $G$ forms a closed chain, we call that a cycle; tree graphs are graphs which have no cycle. A complete graph on $N$ vertices is one where every vertex has an edge connecting it to every other vertex, i.e. it is the unique $(N-1)$-regular graph on $N$ vertices. 

Given an underlying graph $G$, we can define a class of tensor network states, by assigning a set of tensors located on each vertex, where the virtual bonds correspond to the edges connecting that site (see Fig.~\ref{fig:tn_graph}a-f). A tensor network state with uniform bond dimension (i.e. the dimension of the virtual space) $\chi$ and physical dimension $d$ on $r$-regular graph is specified by a set of $r+1$ rank tensors with $d\chi^{r}$ entries. 

Matrix product states form a class of these general tensor network states when the underlying graph is a 1-dimensional lattice (Fig.~\ref{fig:tn_graph}b). Accessing local operators on such 1-d tensor network states is efficient because there is an efficient algorithm to contract it, which depends on the fact that any connected subgraph can be separated from the lattice by cutting only 1 or 2 edges; hence any matrix manipulation during the contraction procedure scales linearly with the number of vertices $N$ of the graph. This feature works not just for 1-d, but for any tree-like geometry, which is why tree tensor networks (Fig.~\ref{fig:tn_graph}c) are efficient ansatz for any tree-like quantum model~\cite{Shi_2006,nakatani2013efficient}.  However this does not hold true for 2-d lattices on $N$ vertices (Fig.~\ref{fig:tn_graph}d). The number of dangling edges of any connected subgraph (or the ``surface area") for such lattices can be upto $\sim \sqrt{N}$, which implies that the slowest step for tensor contraction will require manipulating an array of size $\chi^{\sqrt{N}}$ which has an unfavorable exponential scaling. Another way of seeing why contracting 2-d lattices is difficult is by noting that 2-d lattice has cycles at all scales, while tree tensor networks are acyclic, and 1-d lattice has either no cycle (open boundary condition) or one very long cycle (periodic boundary condition).

Lattices are atypical graphs - which can be understood by considering the expansion property of typical graphs. For any subset of vertices $S \subset V$, we define $\mathcal{E}(S)$ to count the number of dangling edges in $G$ with one extremity in $S$ and the other in $V\setminus S$ ($\mathcal{E}(S)$ captures the notion of ``surface area" of S).  One can formally define an expansion coefficient, which is the minimal ratio of the ``surface area" to the volume for any subgraph of G,

\begin{equation}\label{eqref:expansion}
    h(G) = \min_{S} \{\mathcal{E}(S)/|S| \text{ for } \varphi \neq S\subset V \text{ and } |S|\leq N/2\}.
\end{equation} 
For lattices, $h(G) \to 0$ as $N\to \infty$, i.e. the volume scales faster than the ``surface area". However, generic graphs have positive ``expansion", which can be formalized by considering a probabilistic scheme to construct generic graphs, namely random regular graphs \cite{bollobas_2001}. A random regular graph is a graph drawn from a probability space $\mathcal{G}_{N,r}$, which are all r-regular graphs on $N$ vertices. It can be shown that a random regular graph is an expander graph for large $N$ with high probability, i.e. they have positive expansion, $h(G) = c > 0$ \cite{bollobas_2001}. Note, these graphs are sparse and the number of edges only scales linearly with $N$, since the graph is $r-$regular. 

From this general definition, it would seem that manipulations of tensor networks defined on such expander graphs would be prohibitively inefficient, as a naive contraction will now have to deal with matrix multiplication over an index that scales as badly as $\chi^{\mathcal{O}(N)}$. At the same time, because of the underlying graph structure, a tensor network defined on such a graph can represent a volume law entangled state with even a finite bond dimension $\chi$. The entanglement of a subregion of the tensor network satisfies, $S\leq cN \log \chi $, where $c$ is the expansion, and hence the maximal entanglement scales as a volume-law. For $\chi >  e^{\log{2}/c}$, the above tensor network bound exceeds the universal bound $S(A) \leq |A| \log 2$. This suggests that with $\chi \sim e^{\log{2}/c}$ but independent of $N$ one can already represent nearly maximally entangled state on asymptotically large contiguous subsystems. Hence, we expect that highly entangled states on such graphs can be represented with very modest bond dimension.

However, generic expander graphs are also locally tree-like, which arises due to their sparsity. Typical graphs of $\mathcal{G}_{N,r}$ have a small number of short cycles. In fact, asymptotically $N\to \infty$, the number of cycles of length $i$ behaves as independent Poisson random variable with mean $(r-1)^{i}/(2i)$. Note, however that the diameter of an expander graph is $\sim \log N$, so for a given sized graph, the cycles of size $\log N$ must exist. Still, starting from any vertex, at large $N$ one has to go farther and farther to see any cycles at all, which makes these graphs `locally' tree-like.

This locally tree-like feature is special to expander graphs, which is not present for d-dimensional lattices. We find that this property actually allows us to contract tensor networks defined on such graphs efficiently.



\section{Belief propagation method to contract tensors}\label{sec:bp}

\begin{figure}
    \centering
    \includegraphics[width=\columnwidth]{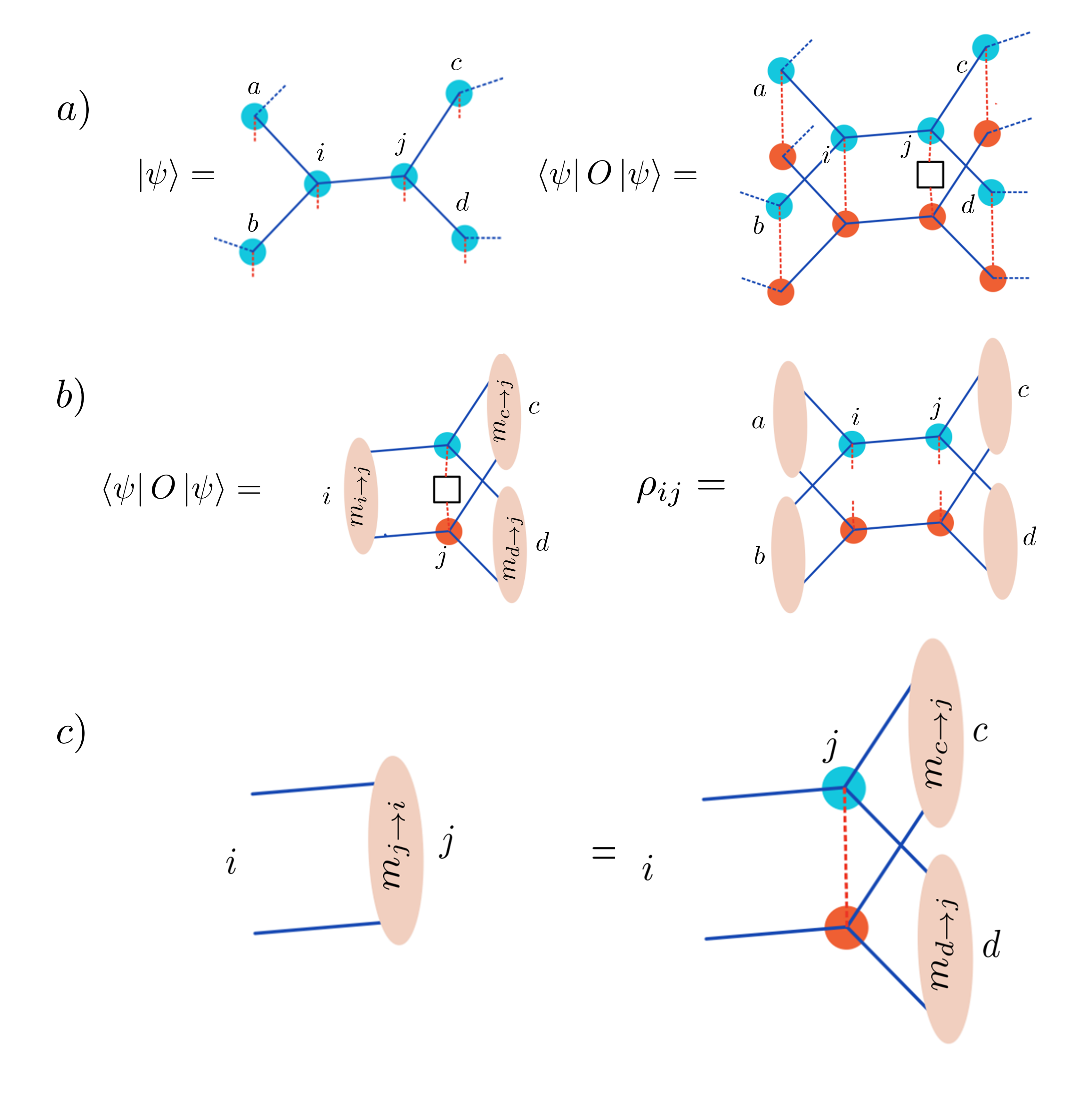}
    \caption{\textbf{Belief propagation algorithm to contract tensor network on a graph.} \textbf{(a)} shows a patch of the tensor network state $\ket{\psi}$. Expectation value of any local operator $O$ can be computed by considering $\ket{\psi}$ and its conjugated copy and contracting them. These can be equivalently computed using the message tensors, as shown in \textbf{(b)}. The reduced density matrix $\rho_{ij}$ of the state $\ket{\psi}$ for two nearest neighbor sites in terms of the local message tensors is shown as well.
    \textbf{(c)} pictorially depicts the central BP equation Eq.~\ref{eqref:msg_tns_1}, which is iterated (as in Eq.\ref{eqref:bp_tns}) to find fixed points of the message tensors.}
    \label{fig:bp_tns}
\end{figure}
We now describe the belief propagation algorithm for contracting tensor networks, following the method introduced in~\cite{Alkabetz_2021}. Suppose we are given a tensor network state $\ket{\psi}$ defined on an underlying graph $G$. Computing the norm $\bra{\psi}\psi\rangle$ or expectation value of an operator $\bra{\psi}\mathcal{O}\ket{\psi}$ requires us to take two copies of the tensor network (one with a complex conjugate), stack them and introduce the operator $\mathcal{O}$ if necessary, and trace over the physical legs (see Fig.~\ref{fig:bp_tns}a). The resulting network is a double-edged factor graph~\cite{Alkabetz_2021}, and we will use BP to compute its marginals, which in our case corresponds to the local reduced density matrices. 

We define a `message' tensor $m_{a\to b}$ corresponding to each directed edge connecting two vertices $a,b\in G$, with an added direction $a\to b$. $m_{a\to b}(x,x^{\prime})$ is a $\chi \times \chi$ dimensional tensor which corresponds to the contraction of the tensor and its conjugate for all sites in $G$ which are connected to $b$ via $a$. In $m_{a\to b}(x,x^{\prime})$, $(x,x^{\prime})$ refer to the indices corresponding to the virtual bonds along $ab$ of the tensor $\ket{\psi}$ and its conjugate, which also makes the matrix $m_{a\to b}$ positive semi-definite. By this definition, one can set up a recursive self-consistency relation that relates the message tensor to their nearest neighbor state tensor $\psi_{i}$ and the next-nearest neighbor message tensors, 
\begin{equation}\label{eqref:msg_tns_1} 
m_{i\to j} = \Tr\left(\psi_{i}\psi_{i}^{*}\prod_{k \in \mathcal{N}_{i}\setminus j}m_{k\to i}\right).
\end{equation} Here, $\mathcal{N}_{i}\setminus j$ refers to the neighboring vertices of $i$ apart from $j$. The self consistency Eq.~\ref{eqref:msg_tns_1} can be pictorially represented as shown in Fig.~\ref{fig:bp_tns}c.

Note, the definition of $m_{a\to b}$ as the result of the contraction of the tensor network connected to $b$ `via'  $a$ only makes sense when the underlying graph has no closed chain connecting $a$ and $b$, i.e. the graph is a tree. However, the recursive definition Eq.~\ref{eqref:msg_tns_1} is a consistent definition for a positive definite message tensor that works for any graph. Our goal is to access self-consistent message tensors that satisfy the recursive equation Eq.~\ref{eqref:msg_tns_1} by the Belief Propagation algorithm, and then identify that as the result of an actual contraction of the tensor network. In order to do that, we simply iterate the self consistency equation, starting from some initial choice of positive semi-definite message tensors for each directed edge of the graph at $t = 0$. At any subsequent time-step, we get message tensors,
\begin{equation}\label{eqref:bp_tns}
m_{i\to j}^{[t+1]} = Tr\left(\psi_{i}\psi_{i}^{*}\prod_{k \in \mathcal{N}_{i}\setminus j}m^{[t]}_{k\to i}\right).
\end{equation}
We look for fixed points of this iterative algorithm. Note that this can be an uncontrolled step, and in general we are neither guaranteed that a fixed point exists, nor that the fixed point corresponds to the correct marginal contraction. Furthermore, there is a `gauge' freedom in the definition of the message tensor, as many message tensors can correspond to the contraction of the same tensor network state. However, for tree-like graphs, this is guaranteed to converge to the result from the contracted tensor. As was pointed by~\cite{Alkabetz_2021}, even on a 2-d lattice where there is a proliferation of short cycles, this algorithm can return good approximate answers.

Extracting local reduced density matrices is straightforward once we have the self-consistent message tensors, and only requires local contraction of the state tensor with the message tensor, as shown for a neighboring 2-site reduced density matrix in Fig.~\ref{fig:bp_tns}b.

The central BP equation for tensor network contraction Eq.~\ref{eqref:bp_tns}, for an underlying graph with cycles is a version of the loopy-BP algorithm. As mentioned in the introduction, while loopy-BP is not guaranteed to succeed, it has been shown to work extremely well in many practical scenario. Perhaps its most useful application lies in the decoding of low-density-parity-check (LDPC) codes~\cite{McEliece_1998}. Importantly, LDPC codes are asymptotically locally-tree like, hence the effective `Bethe' or tree-approximation inherent in the BP algorithm works well there.

Motivated by this observation, we employ the BP algorithm to access local expectation values for tensor networks defined on random regular graphs. The intuition is as follows: consider a state on the graph $G$ with correlation length $\xi$. Typical lengths of cycles on random regular (and in general sparse expander) graphs is $\sim \log N$. Hence, if $\xi < \log N$, the state is expected to look tree-like, and the BP algorithm should converge to the right answer. Crucially, in sparse expander graphs the typical cycle length diverges in the thermodynamic limit, so one can expect successful contraction of a wide scale of states which are not just short-range correlated.

\section{Graph states and square root states}\label{sec:graph_sqroot}
In this section we demonstrate the usefulness of this method to extract local information from a class of tensor network states defined on sparse graphs. We introduce a class of tensor network states for qudits with local Hilbert space dimension $d$ which can be efficiently represented as a tensor network with bond dimension $\chi = d$ on any underlying graph $G$,
\begin{equation}\label{eqref:gen_graph_state}
    \ket{\psi} \sim \sum_{\{s\}} \left(\prod_{ab\in \text{Edge}_G}M(s_{a},s_{b})\right)\ket{\{s\}},
\end{equation}
where $s_{a}$ is a basis of the d-dimensional local qudit Hilbert space. These states can be called generalized graph states. We consider the following decomposition of the $d\times d$ dimensional matrix $M = A A^{T}$. Now, the state in Eq.~\ref{eqref:gen_graph_state} can be constructed out of the $A$ matrices explicitly and locally. Consider a vertex $a\in G$, with degree $r$. Consider the generalized identity tensor $I_{s,\alpha_{1},..,\alpha_{r}} = \delta_{s\alpha_{1}}\delta_{2}...\delta_{s\alpha_{r}}$, where the $s$ index refers to the physical qudit index and $\alpha_{i}$ refer to the $r$ virtual indices. Now we can multiply the $A$ matrices to $I$ to get the local tensors $T$ corresponding to the state in Eq.~\ref{eqref:gen_graph_state}, 
\begin{equation}
    \includegraphics[width = 0.7\columnwidth]{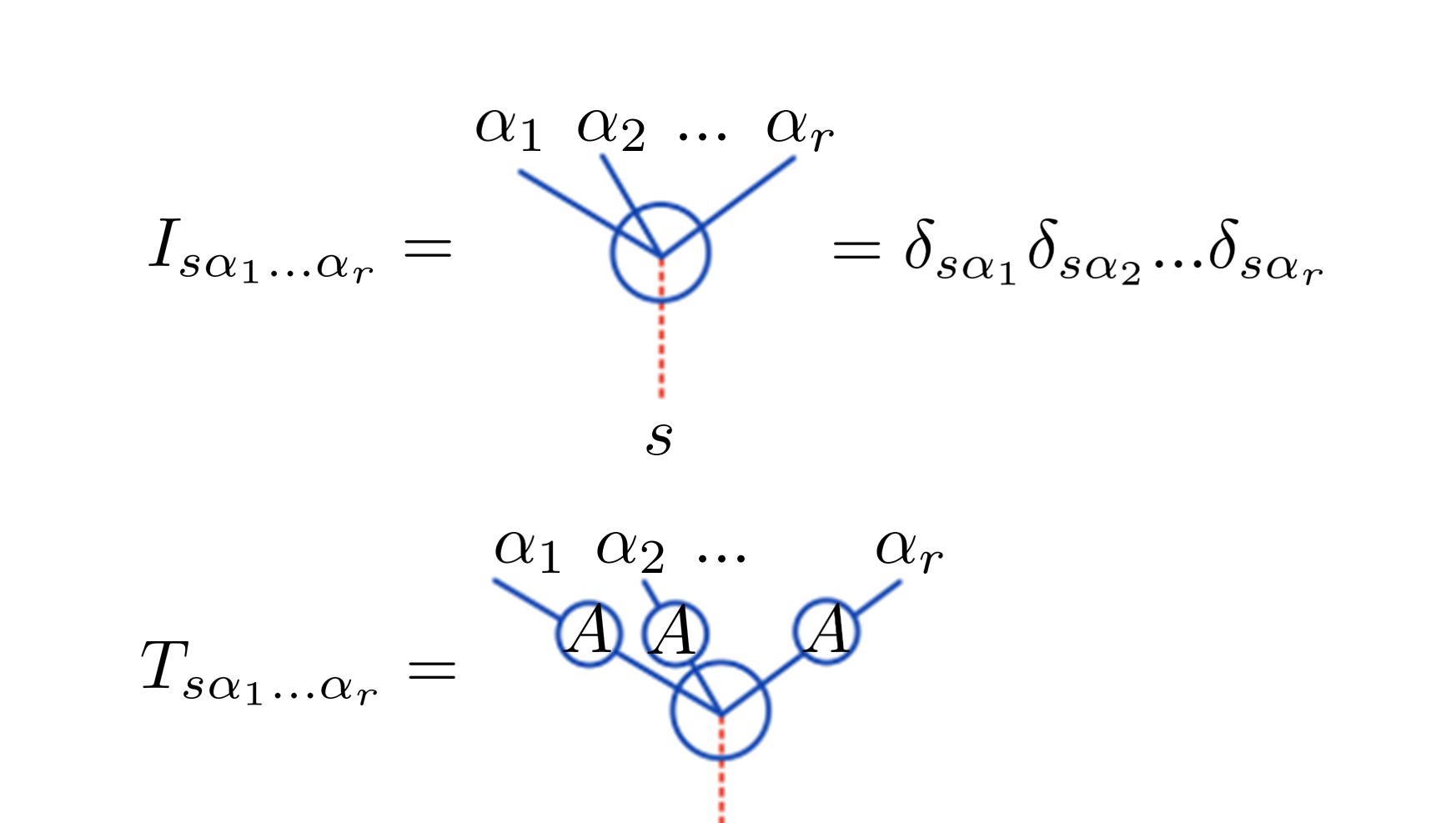}
\end{equation}

\subsection{Square root states of classical models}
Consider a classical Ising model on any generic graph,
\begin{equation}
    H_{c} = - J \sum_{i, j\in \mathcal{N}_{i}}s_{i}s_{j},
\end{equation}
with the partition function $\mathcal{Z}(\beta) = \sum_{\{s\}}e^{-\beta H_{c}}$.

We consider the square root state associated with it \cite{Swingle2016},
\begin{equation}\label{eqref:sq_state_def}
    \ket{\psi} = \frac{1}{\sqrt{\mathcal{Z}(\beta)}}\sum_{\{s\}}e^{+\frac{\beta J}{2}\sum_{i,j\in \mathcal{N}_{i}}s_{i}s_{j}}\ket{\{s\}}.
\end{equation}
We denote the Pauli spin operators as $X,Y,Z$ and the identity operator as $\mathbf{1}$. These states are called square root states, since these can be understood to be the square root of the Ising model partition function; in fact if we consider the unnormalized state, $\tilde{\ket{\psi}}= \sum_{\{s\}}e^{+\frac{\beta J}{2}\sum_{i,j\in \mathcal{N}_{i}}s_{i}s_{j}}\ket{\{s\}}$, the classical partition function is equal to its norm $\mathcal{Z}(\beta) = \tilde{\bra{\psi}}\tilde{\psi\rangle}$. Expectation value of any classical operator (i.e. an operator constructed out of $Z_{i}$ operators) in the state $\ket{\psi}$ is equal to an averaged classical statistical quantity,
\begin{equation}
    \bra{\psi}Z_{a}\ket{\psi} = \langle Z_{a}\rangle_{H_{c}} = \frac{\Tr_{\{s_{i}\}} Z_{i}e^{-\beta H_{c}(\{s_{i}\})}}{\Tr_{\{s_{i}\}}e^{-\beta H_{c}(\{s_{i}\})}}.\label{eq:classical_quantum}
\end{equation}
The latter can be estimated by classical Monte Carlo methods, so we can access the expectation value of classical operators easily. However, accessing quantum operators, for example $\langle X_{a}\rangle$ is not possible using a naive Monte Carlo approach. 

\begin{figure}
    \centering
    \includegraphics[width=\columnwidth]{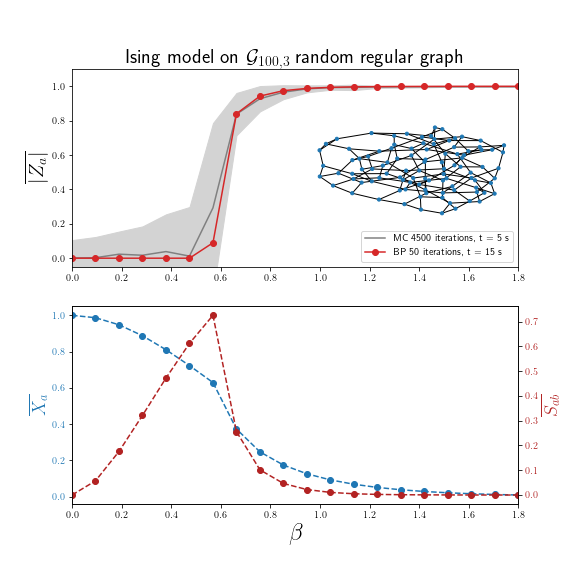}
    \caption{\textbf{Expectation values of local operators of Ising model square root states}, defined in Eq.~\ref{eqref:sq_state_def}, on a random regular graph from the ensemble $\mathcal{G}_{100,3}$ [inset]. $J$ is set to be $1$. In the \textbf{top} panel, absolute value of local $Z$ operator averaged over the vertices of the graph is plotted as a function of $\beta$. Since this is a classical observable, it can be estimated by straight-forward Monte Carlo sampling (MC), which is shown with the error bar estimate from the average. The BP result is shown in red, which matches the MC estimate. In the  \textbf{bottom} panel we show the BP result of the site-averaged $X$ operator and the edge-averaged entanglement entropy of reduced density matrix of nearest neighbor sites. These expectation values are inaccessible to simple MC sampling of the classical model.}
    \label{fig:sq_root_bp_mc}
\end{figure}

The square root state in Eq.~\ref{eqref:sq_state_def} can be shown to be the ground state of a parent quantum Hamiltonian defined on the graph,
\begin{equation}
    H = \sum_{a}\left[-X_{a}+e^{-\beta J Z_{a}\sum_{b\in \mathcal{N}_{a}}Z_{b}}\right].
\end{equation}
It also corresponds to a general graph state defined in Eq.~\ref{eqref:gen_graph_state}, with $\chi = d = 2$,
\begin{equation}
    M(s_{a},s_{b}) = \begin{pmatrix}
\exp{\beta J/2} & \exp{-\beta J/2}\\
\exp{-\beta J/2} & \exp{\beta J/2}
\end{pmatrix}
\end{equation}
and the corresponding $A$ defined by $AA^{T} = M$ can be computed straight-forwardly. We study the expectation values of `classical' $Z$ and the `quantum' $X$ operators averaged over all sites of the graph using the BP contraction method. The results are shown in Fig.~\ref{fig:sq_root_bp_mc}, where the underlying graph is taken to be an instance of random regular graph $\mathcal{G}_{N=100,r=3}$.

We plot $\overline{|Z_{a}|}$, which is the absolute value of the expectation of $Z_{a}$ operators averaged over all vertices of the graph, as a function of the inverse temperature $\beta$ that is a parameter in the theory. Since this quantity can be directly computed using Monte-Carlo sampling on the original Ising model, we get an independent check for the BP method. We find that $\overline{|Z_{a}|}$ is an order parameter for the phase transition in the classical model that occurs from the paramagnetic phase at low $\beta$ to an ordered ferromagnetic phase at high $\beta$, and the results are consistent between the BP and the MC answers across all $\beta$. However, using the BP messages, we can also compute the averaged $\overline{X_{a}}$ expectation values, and the averaged entanglement of 2-site reduced density matrices on all edges $ab \in G$. Note, these quantities are not easily accessible via naive Monte Carlo sampling of the classical model, which shows an application of the BP method of tensor network contraction. We discuss the convergence issues and the `gauge' freedom of the BP message tensors in Appendix~\ref{appsec:graph_sq}.

\subsection{Graph states on sparse graphs}
Graph states in quantum computing ~\cite{Hein_2006} are generalizations of cluster states which are resources for measurement-based quantum computing ~\cite{Briegel_2009}. Importantly for this work, graph states are a kind of generalized graph state as defined in Eq.~\ref{eqref:gen_graph_state}.

\begin{figure}
    \centering
    \includegraphics[width = \columnwidth]{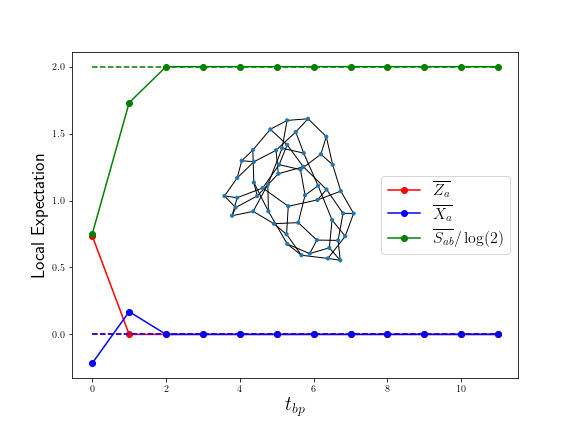}
    \caption{\textbf{Local expectation values of graph states} defined on a $\mathcal{G}_{50,3}$ random regular graph, as a function of the number of BP steps. The BP steps converge to the correct expectation value for the 1-body and 2-body expectation values after 3 steps.}
    \label{fig:graph_state}
\end{figure}

Graph state $\ket{G}$ is pure state on $N$ qubits for a graph $G$ on $N$ vertices. We start with the product state $\left(\ket{0}+\ket{1}\right)^{\otimes N}/2^{N/2}$, and apply the controlled phase gate $U = \ket{0}\bra{0}\otimes \mathbf{I}+\ket{1}\bra{1}\otimes Z$ to any pair of qubits on vertices connected by the edges in $G$. It can be easily shown that in our definition of generalized graph states in Eq.~\ref{eqref:gen_graph_state}, $\ket{G}$ corresponds to,
\begin{equation}
    \label{eqref:graph_states}
    \ket{G} \equiv M(s_{a},s_{b}) = \begin{pmatrix}
    1 &1\\1&-1
    \end{pmatrix}.
\end{equation}
Hence $\ket{G}$ is efficiently represented by a $\chi = 2$ bond dimension tensor network state defined on any graph $G$. Graph states form an important class of multi-party entangled states.

We can estimate the local expectation values of $\ket{G}$ defined on random regular graphs using the BP algorithm. On a $3$-regular graph, the expectation value of any of the local Pauli operators $X,Y,Z$ is $0$, and the entanglement of any reduced density matrix on an edge is $2\log 2$, which is confirmed as the fixed point after $\sim 3$ BP steps, as shown in Fig.~\ref{fig:graph_state}. 

These results demonstrate the utility of the BP algorithm in accessing local expectation values of a class of entangled states defined on graphs. Next, we introduce a variational algorithm which uses the BP algorithm as a subroutine, to approximately determine the ground state energy and prepare an approximate ground state of a quantum model defined on a sparse graph.
 
\section{Variational preparation of ground states of sparse graph Hamiltonians}\label{sec:variational_prep}
Suppose we are given a graph G and a Hamiltonian defined on it, $H = -\sum_{a,b\in \mathcal{N}_{a}}h_{ab}$. Our goal is access the ground state $\ket{\psi_{GS}}$ of this $H$, and estimate its energy, $E_{GS}$. Note, given a tensor network state $\ket{\psi}$, the estimation of its energy can be achieved by computing energy functionals over 2-body local reduced density matrices along the edges, $\bra{\psi}H\ket{\psi} = \sum_{ab}\bra{\psi}h_{ab}\ket{\psi}$, which can be estimated using the BP algorithm. This suggests a variational method to prepare an approximate ground state.

We first start with an initial state $\ket{\psi_{in}}$. At each variational step, we perform a fixed number of BP iterations to access the approximate message tensors. Next, we fix the message tensors $\{m\}$, and locally update the state tensor $\ket{\psi}$ by gradient descent to minimize the energy functional, $\ket{\psi} \to \ket{\psi}-\alpha \nabla H\left(\ket{\psi},\{m\}\right)$, where the gradient of the energy functional is computed using fixed messages $\{m\}$ obtained beforehand. These two steps are repeated until the energy of the state reaches a steady value. The pseudocode is provided here, 

\begin{algorithm}[H]
\caption{Variational TN ground state preparation}\label{algo:variational}
\begin{algorithmic}
\State Initialize state $\ket{\psi} =\ket{\psi_{in}}$
\While{$t < t_{var}$}
    \While{$\tau < t_{bp}$}
        \State BP on $\ket{\psi}$: messages $\{m[\tau+1]\} =  BP_{\ket{\psi}}(\{m[\tau]\})$
    \EndWhile
    \While{$n < n_{gd}$}
        \State Gradient descent: $\ket{\psi} \to \ket{\psi} - \gamma \nabla H\left(\ket{\psi},\{m[t_{bp}]\}\right)$
    \EndWhile
\EndWhile
\end{algorithmic}
\end{algorithm}

A comment on the variational method: it does not guarantee physically realistic local expectation values during the variational steps. The quantum states are always tensor network states and hence physical, however the BP steps are iterated for a prefixed finite time and not until they have converged on to the messages corresponding to the tensor network states.

As a demonstration we consider a random regular graph drawn from $\mathcal{G}_{N,r=3}$, and define a nearest neighbor mixed-field Ising Hamiltonian, with edge terms, $h_{ab} = Z_{a}Z_{b}$ and vertex terms, $h_{a} = 2 X_{a} + 0.5Z_{a}$. We consider the following parameters for the variational procedure, $t_{bp} = 5, n_{gd} = 10, \gamma = 0.01$, and the initial states are chosen to be either product states or high temperature square root states of the classical Ising Hamiltonian Eq.~\ref{eqref:sq_state_def}.

\begin{figure*}
    \centering
    \includegraphics[width = \textwidth]{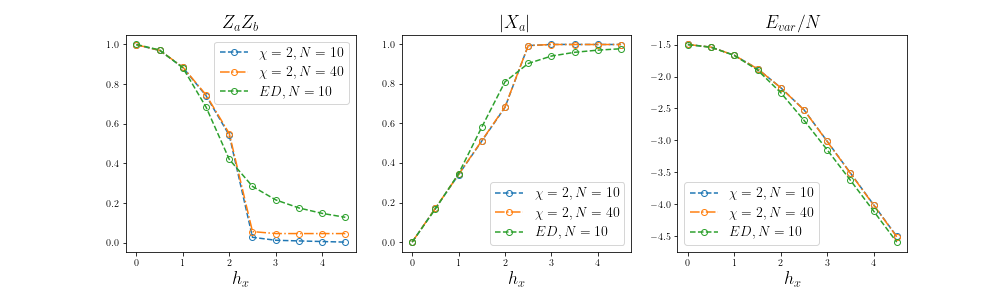}
    \caption{\textbf{Quantum Ising model on a random regular graph}. We variationally access the local order parameters and the energy density for both $N = 10$ and $N=40$ sized random regular graphs $\mathcal{G}_{N,r=3}$, with tensor network states with $\chi = 2$. The results are also compared with the $N = 10$ exact diagonalization data.}
    \label{fig:quantum_ising_model}
\end{figure*}

The results are shown in Fig.~\ref{fig:variational_gs_prep_benchmark} in Sec.~\ref{sec:intro}, which is a benchmark study for $N = 10$ for which the exact ground state can obtained by exact diagonalization. We plot the estimated ground state energy, and the fidelity of the obtained state with the exact ground state for two different bond dimensions $\chi = 1,2$. We find that with $\chi = 2$ the energy of the variational state is indistinguishable from the exact ground state energy, and the overlap with the exact ground state also is significantly higher than any random state. This suggests that we have variationally prepared a $\chi = 2$ tensor network state which is very close to the exact ground state of the Hamiltonian. This method can be readily generalized to $N = 40$ which takes $< 20$ minutes on a standard 16 GB laptop to run (see results in Fig.~\ref{fig:variational_gs_prep}); however these sizes are inaccessible to exact diagonalization. 

Variational ground state preparation on quantum graphical models using quantum belief propagation has been studied before~\cite{Ramezanpour_2012,Biazzo_2013}. On the other hand, variational tensor network ground states have also been studied on tree lattices~\cite{Nagaj_2008,Lunts_2021}, where the tensor network contraction is simple because of the lack of cycles or loops. Using the formulation of BP for tensor network states, our method leads to approximate tensor network representation of the whole ground state, from which correlation functions may be estimated.

As another demonstration, we consider the quantum Ising model with transverse field,
\begin{equation}
    H = -\sum_{ab}\left(Z_{a}Z_{b} +h_{x}X_{a}\right),
\end{equation}
and access local expectation values of $Z_{a}Z_{b}$, $X_{a}$, and the energy density, as a function of $h_{x}$, shown in Fig.~\ref{fig:quantum_ising_model}. This model undergoes the standard $\mathbb{Z}_{2}$ symmetry breaking quantum phase transition. We find that the variational method works well in the gapped ferromagnetic ($h_{x} \ll 1$) and the paramagnetic ($h_{x} \gg 1$) phases, but deviates from the finite size exact diagonalization data near the transition. Interestingly, the local order parameters and the energy density accessed using the BP method show the same results for both the $N=10$ and $N=40$ sized graphs return similar values. This indicates that the BP method is able to access the local properties of the large-$N$ graph even with the small finite sized numerics. The results indicate there is a phase transition at $2\leq h_x\leq3$. In the ferromagnetic phase of the transverse field Ising model the variational method produces a state in the `degenerate' ground space, which is in general an uncontrolled superposition of the two lowest lying states; however the local order parameters do not distinguish between the states (see discussion in Appendix~\ref{appsec:qIsing}). 

\begin{figure}
    \centering
    \includegraphics[width = \columnwidth]{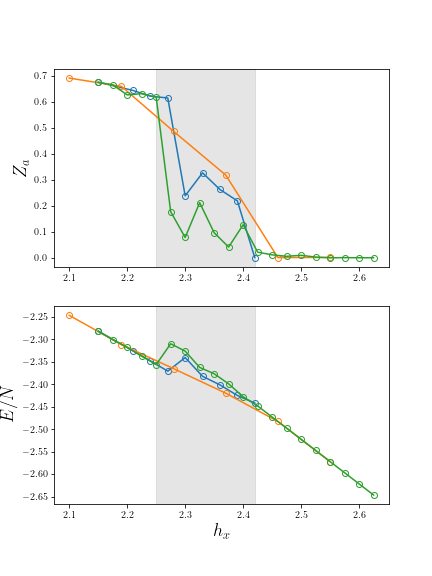}
    \caption{\textbf{Quantum Ising model near criticality.} We consider the transverse field quantum Ising model on a random regular graph $\mathcal{G}_{40,3}$, and plot the local order parameter $Z_{a}$ averaged over all sites, and the energy density as a function of the field $h_{x}$. The different traces are different runs of the variational algorithm, starting with slightly different initial states, and running for a constant number of iterations which converge away from the critical point.}
    \label{fig:qIsing_criticality}
\end{figure}

In Fig.~\ref{fig:qIsing_criticality} we zoom into the critical region, and access the local order parameter $Z_{a}$ (averaged over all sites) and the energy density. In \cite{Nagaj_2008}, this model was studied on Bethe lattices using imaginary time evolution, and the phase transition was characterized to be mean-field like. In the random regular case, we find that the variational method slows down considerably near the critical point, and we are not able to access consistent results after a finite number of iterations ($t_{var} = 15$) when we start with distinct initial states. Fig.~\ref{fig:qIsing_criticality} indicates that the phase transition occurs at $2.25\leq h_{x}\leq2.45$, but we are not able to characterize the critical properties of the transition using the variational method for tensor networks with $\chi = 2$. 

In a Bethe lattice with degree 3, which is locally similar to the random regular graph $\mathcal{G}_{N,3}$ at large $N$, bond dimension 2 is enough to asymptotically represent maximally entangled states on large enough subregions. However, the variational method we have studied is a local update method, which is presumably why it fails to approximate the long-range correlated state near the critical point. The sparse graph model is mean-field like, which implies that the physics near the critical point is governed by the uniform spatial zero-mode. One can consider a uniform ansatz for the variational update which will work better for accessing the critical properties of the transition. Also, one can use better variational methods to tackle the issue of small gaps and slow convergence, such as stochastic versions of gradient descent or simulated annealing.



\section{Discussion}
In this work we have demonstrated that tensor networks on generic graphs can be contracted using the belief propagation algorithm, and these work very well for studying quantum systems defined on locally tree-like graphs. We demonstrated the usefulness of such a method by using it extract local information of tensor network states defined on such graphs, such as graph states and square root state of classical Ising model. We also developed a variational method to prepare an approximate ground state of a gapped quantum spin model defined on random regular graphs. We then used this method to also study the phase diagram of the quantum Ising model with transverse fields.

These results open up several new avenues of research. Firstly, there may be application of more developed BP algorithms~\cite{Yoon_2011,Kirkley_2021} to study tensor network states on graphs with short loops and 2-d tensor network states with short correlation length. In particular, these modified BP algorithms can tackle short loops efficiently: in gapped local systems one can systematically increase the maximal short loop size until it crosses the correlation length, and obtain accurate contraction of PEPS tensor network states. Systematic study of the limitation of the BP algorithm in contracting tensor network states might also shed light on the exotic nature of the states, for e.g. spin glass order. 

Another direction of study would be to study time evolution of sparse graph tensor network states under a Hamiltonian defined on the graph, by using BP in tandem with algorithms such as the density matrix renormalization group (DMRG). These will be useful for studying both dynamics as well as accessing the ground state via imaginary time evolution. One restriction to this is the process of mid-circuit truncation of the graph tensor network states, which is not guaranteed to be an appropriate truncation of entanglement when there are loops. One direction of approach would be to use the BP algorithm for efficient truncation of graph tensor network states~\cite{Evenbly_2018}, which would be an essential step towards accurate DMRG on such graphs. At the same time, even naive truncation of the tensor network may already be sufficient for sparse graphical models, as they are locally-tree like. On a related note, studying quantum many-body systems on such expander graphs may lead to new physical insights about the nature of the nature of the many-body groundstates and its associated quantum error correction, inspired by novel quantum error-correcting LDPC codes which are under intense recent study in the quantum information community~\cite{Breuckmann_2021}.

Simulating real time evolution using tensor networks on lattice systems is generally limited by the entanglement and the bond dimension. However, as mentioned in Sec.~\ref{sec:graph_tn}, the graph structure of `expander' graphs allow for an efficient representation of highly entangled states with only modest bond dimensions. Hence, one should in principle be able to track entanglement build-up for long times with only polynomial resources. This is also a promising direction of future studies. 



Another interesting question would be to explore tensor network states associated with multi-body spin and fermionic Hamiltonians on graphs. The variational method for accessing the ground state lends itself naturally to Hamiltonians with multi-body terms. On the other hand, one can also set up a fermionic tensor network states by using parity symmetric tensors and fermionic SWAP gates~\cite{Corboz_2010}. This suggests a pathway towards simulating interacting fermionic models on sparse graphs.





\section{Acknowledgement}
We acknowledge useful conversations with and feedback from Chris Baldwin, Christopher White, and Shenglong Xu. The work of SS is partially
supported by the U.S. Department of Energy, Office of Science, Office of Advanced Scientific
Computing Research, Accelerated Research for Quantum Computing program ``FAR-QC''. The work of BGS is supported in part by the AFOSR under grant number FA9550-19-1-0360.

%

\newpage
\begin{appendix}
\section{Details of the numerical implementation}
\label{appsec:numerics}
In this section we provide details of some of the numerical observations behind the results in the main paper. We will also discuss convergence and related issues.
\begin{figure}[ht!]
    \centering
    \includegraphics[width = 0.9\columnwidth]{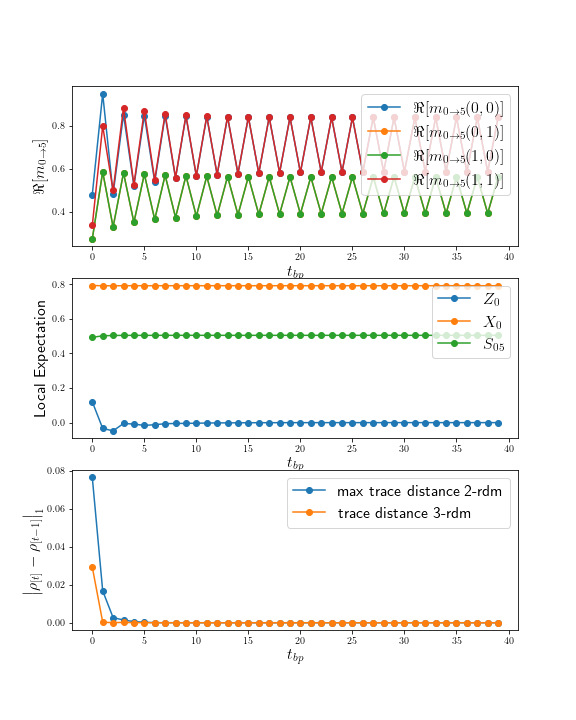}
    \caption{Convergence of BP messages and local expectation values with BP for Square root state with $\beta = 0.4 J^{-1}$ on a $\mathcal{G}_{20,3}$ random regular graph.}
    \label{fig:bp_convergence_sqroot}
    \centering
    \includegraphics[width = 0.9\columnwidth]{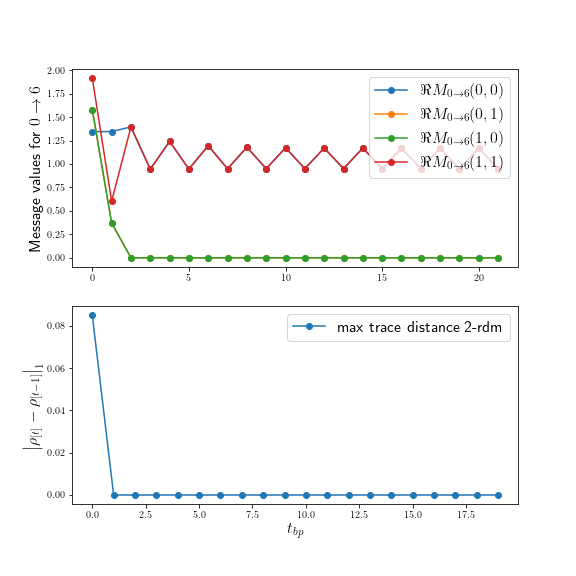}
    \caption{Convergence of BP messages and local expectation values with BP for a graph state on a $\mathcal{G}_{20,3}$ random regular graph.}
    \label{fig:graph_state_appendix}
\end{figure}

\subsection{Graph states and Square root states}\label{appsec:graph_sq}
Here we show the convergence results for the message tensors and local expectation values of the square root states of classical Ising model (Eq.~\ref{eqref:sq_state_def}) and the graph states (Eq.~\ref{eqref:graph_states}), both of which are exact tensor network states with $\chi = 2$. 

In Fig.~\ref{fig:bp_convergence_sqroot} we study the square root state at $\beta = 0.4 J^{-1}$. In the top panel of Fig.~\ref{fig:bp_convergence_sqroot} we plot the message tensors $m_{a\to b}$ (we show only the real part) of a particular directed edge in the graph, with the iterations of the BP algorithm. It is evident that the entries of the message tensors settle into a limit cycle after a few rounds of BP. In the middle panel of Fig.~\ref{fig:bp_convergence_sqroot}, we plot the expectation value of local operators ($X_{a}$ and $Z_{b}$, on the site connected by the edge under consideration), and the entanglement entropy $S_{ab}$ of the edge. In the lower panel of Fig.~\ref{fig:bp_convergence_sqroot}, we plot the the trace distance between subsequent BP iterations of the reduced density matrices of any 2 and 3 body continuous subregions. For the 2-body reduced density matrix we consider the maximal value of the trace distance over all edges, while for the 3-body reduced density matrix we only consider a particular set of 3 connected neighbors). The middle and lower panels show that the local expectation values have converged, and the reduced density matrices have converged in trace distance. The limit cycle of the message tensor indicates that there exist a notion of `gauge' equivalence between different message tensors which lead to the same expectation values.

Similar feature can be seen in the convergence study for graph states on a random regular graph $\mathcal{G}_{40,3}$, as shown in Fig.~\ref{fig:graph_state_appendix}. Here also the entries of the message tensor show a limit cycle behavior which also coincides with a convergence of reduced density matrices. 

\begin{figure}
    \centering
    \includegraphics[width = 0.8\columnwidth]{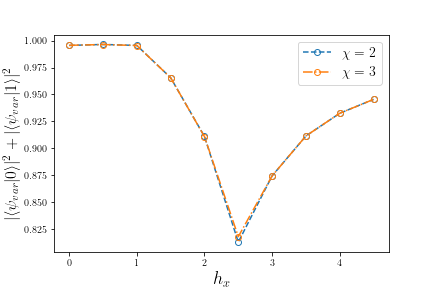}
    \caption{Fidelity of variational state in transverse field Ising model on a random regular graph $\mathcal{G}_{20,3}$. The overlap of the variational tensor network wavefunction $\psi_{var}$ with $\chi = 2$ and the two lowest energy states (which are accessed by exact diagonalization) is shown as a function of the transverse field $h_{x}$.}
    \label{fig:quantum_ising_fidelity}
\end{figure}

\subsection{Variational ground state of the transverse field Ising model}\label{appsec:qIsing}
Here we show details of the fidelity of variational ground state preparation for the transverse field quantum Ising model on a random regular graph. In Fig.~\ref{fig:quantum_ising_fidelity} we plot the overlap of the variational wavefunction $\psi_{var}$ with the two lowest energy states (which are accessed by exact diagonalization) is shown as a function of the transverse field $h_{x}$. The reason we choose the first two low energy states is because in the ferromagnetic phase they are separated by a very small gap, and the variational method in general creates an uncontrolled superposition of the two `degenerate' ground states. We find that the variational method projects to the ground space very effectively in the ferromagnetic and asymptotically in the paramagnetic phase, while showing a cusp near the transition. This also indicates that the $\chi = 2$ ansatz is not sufficient near the critical point, which is associated with long-range critical correlations.

\end{appendix}

\end{document}